\documentclass[12pt]{article}
\pdfoutput=1
\usepackage[a4paper]{geometry}
\usepackage{jheppub, amsmath,amssymb,amsfonts,amsxtra, mathrsfs, makeidx,graphics,graphicx,amsthm,epsfig, ytableau,bm,longtable,float, color,tikz,mathtools,xfrac,footnote,rotating, lscape, makecell, environ,mathtools, empheq}

\usepackage{epsfig,amssymb} 
\usepackage{amsmath}

\usepackage{amscd,color}
\usepackage{amsmath,graphicx}
\usepackage{graphicx}
\usepackage{verbatim}
\usepackage{latexsym}
\usepackage{amsmath,amsfonts,amssymb,amsthm}
\usepackage{amsmath,amsthm}

\def\be{\begin{equation}}
\def\ee{\end{equation}}

\title{
\begin{center}
Lens space index and global properties \\ for 4d $\mathcal{N}=2$ models
\end{center}}

\author[a]{Antonio Amariti,} 
\author[b]{Andrea Marcassoli} 
\affiliation[a]{INFN, Sezione di Milano, Via Celoria 16, I-20133 Milano, Italy}
\affiliation[b]{Dipartimento di Fisica, Universit\`a di Milano, \\  Via Celoria 16, I-20133 Milano, Italy}

\emailAdd{antonio.amariti@mi.infn.it,andrea.marcassoli93@gmail.com}
\abstract{
The additional data necessary to univocally fix the gauge group for a given algebra are represented by the same charge lattices of mutually local Wilson and 't Hooft lines 
for both 4d  $\mathcal{N}=4$ SYM and  $\mathcal{N}=2$ elliptic models. Motivated by this equivalence in this paper we study the Lens space index of these $\mathcal{N}=2$ elliptic models.
The index is indeed sensitive to the global properties and in the $\mathcal{N}=4$ case it is expected to coincide among S-dual models with different global properties, while it gives different results
for models that lie in other S-duality orbits. Here by an explicit calculation we show that the same results hold for the $\mathcal{N}=2$ elliptic models as well.
}

\begin{document}
\maketitle


\section{Introduction}

S-duality is an intriguing symmetry, originally conjectured by \cite{Montonen:1977sn} for the case of 4d
$\mathcal{N}=4$ SYM.
When the gauge group is $SU(N)$ S-duality acts by inversion on the holomorphic gauge coupling, $\tau \rightarrow -\frac{1}{\tau}$  providing a strong/weak duality.
When this inversion is considered together with another symmetry acting on the gauge coupling, $\tau \rightarrow \tau+1$, the S-duality 
group is $SL(2,\mathbb{Z})$. This provides an origin of this symmetry in terms of the invariance of the dilaton of 10 SUGRA,
connecting the S-duality group of $\mathcal{N}=4$ SYM to the S-duality of string theory.
S-duality has been further extended to cases with lower supersymmetry (see e.g. \cite{Witten:1997sc}). 
For example, when considering elliptic models, it
has been shown that the full S-duality group is generated by the two generators above and other shift and permutation symmetries.
In this case the group becomes the mapping class group of the punctured Riemann surface that describes the 
M-theory construction of the model.
Myriads of applications and connection with the geometry (\emph{e.g.}  the string embedding, lower dimensional dualities, etc...) made S-duality one of 
the more prominent tools in understanding non-perturbative aspects of supersymmetric QFTs in the last decades.

Here we focus on a result obtained in \cite{Aharony:2013hda}
\footnote{See also \cite{Gaiotto:2010be,Donagi:1995cf}}, relating S-duality to the
global properties of the gauge 
group of $\mathcal{N}=4$ SYM.
Concretely, when one considers a QFT with a gauge algebra $\mathbf{g}$ one should provide also additional informations on the 
full gauge group $\mathbf{G}$. In absence of matter fields, being the vector multiplet in the adjoint representation of the 
gauge group, one can equally choose either the universal covering group $\widetilde{\mathbf{G}}$ or some other group
$\mathbf{G}=\widetilde{\mathbf{G}}/\mathbf{H}$, where $\mathbf{H}$ is a subgroup of the center $\Gamma$.
When matter fields are considered, some of the possible groups $\mathbf{G}$ are forbidden, depending on the representation of the matter fields under the gauge group.
This issue does not arise for $\mathcal{N}=4$ SYM because the matter fields are in the adjoint representation.
In this last case additional data are necessary to specify the global structure of the gauge group.
A necessary and sufficient condition, fixing completely such data, has been provided in \cite{Aharony:2013hda}.
 It has been indeed shown that
specifying the spectrum of charges of mutually local line operators, i.e. Wilson and 't Hooft lines 
\cite{Kapustin:2005py},
univocally fixes the gauge group $\mathbf{G}$.
This result has a deep consequence on S-duality. 
This is because by acting with S-duality on the lattice that identifies the spectrum of charges of the line operators 
one can pass from a group $\mathbf{G}$ to a group $\mathbf{G}$', showing that theories with different gauge groups are usually related by an S-duality transformation.
However it is not always the case because there are lattices that are not connected by S-duality to the others, signaling the 
presence of orbits in the S-duality group.
For example, in the case of $\mathcal{N}=4$ $SU(N)$ SYM, these orbits are present whenever squares appear in the decomposition of $N$ into prime factors (i.e. whenever $N$ is not square free).

This mechanism has been extended to cases with $\mathcal{N}=2$ supersymmetry \cite{Amariti:2016hlj}. 
Also in this case, by organizing the spectrum of 
mutually local line operators into lattices of 
electric and magnetic charges it has been possible to associate 
theories with different global properties to S-duality transformations. 
These lattices follows from the presence of bifundamental fields connecting the nodes of the quiver.
In such cases the matter fields have a trivial charge under the diagonal subgroup of the center 
$\Gamma_{\text{diag}} \equiv \text{diag}(\prod_{i=1}^{n_G} \Gamma_i)$, where $n_G$ represents the number of gauge nodes in the quiver.

In this paper we  corroborate this result by studying and comparing the Lens space index of theories with a different global structure but mapped by an $SL(2,\mathbb{Z})$ transformation.
The Lens space index was originally derived in \cite{Benini:2011nc}. It is a supersymmetric index computed on the product space 
$L(r,1) \times S^1$ where $L(r,1)$ corresponds to the Lens space.
Being not simply connected it is sensitive to the global properties of the gauge groups.
Indeed it takes contributions from two simultaneously
conjugated commuting holonomies, one from
the $S^1$ and one from the $S^1$ inside $L(r,1)$.
If the gauge group is the universal covering $\widetilde{\mathbf{G}}$ then the only commuting
 holonomies contributing to the index, say $g$ and $h$, respect a relation  $[g, h]=1$.
If the gauge group is  $\mathbf{G} = \widetilde{\mathbf{G}}/{\mathbf H}$ the relation can be modified, by considering other elements of the center $\Gamma$ instead of the identity.
These contributions come from the so called almost commuting holonomies, i.e. holonomies that do not commute 
in $\widetilde{\mathbf{G}}$ but commute in $\mathbf{G}$. 
In the case of $\mathcal{N}=4$ SYM this index was computed in \cite{Razamat:2013opa} and it was shown that it coincides among  theories
with different gauge group if these theories  are mapped by an S-duality transformation.
It was also observed that in the case of theories living in different orbits of $SL(2,\mathbb{Z})$ the index does not match as expected.
Here we provide the extension of the result to the models with lower supersymmetry discussed above.

The paper is organized as follows.
In section \ref{secrev1} we review the construction of \cite{Aharony:2013hda} and the additional data
necessary for distinguishing models with different gauge groups.
In section \ref{review2} we review the construction of \cite{Razamat:2013opa} for the Lens space index
and its relations with the global properties of $\mathcal{N}=4$ SYM.
In section \ref{sec:n2LSI} we extend the construction to $\mathcal{N}=2$ elliptic models, providing a general 
formula for the Lens space index in these cases.
We apply this formalism to a simple model, namely an $\mathcal{N}=2$
elliptic model with two $SU(N)$ gauge groups
and we show that for $N=2,3,4$ we recover the expected results, i.e. 
the index coincides for all the models in the same S-duality orbit and in the $N=4$
case we also find a different value of the index for the model corresponding to the new orbit.
In section \ref{conclusions} we conclude.
In appendix \ref{toriclines} we furnish the structure of the Lens space index for a generic $\mathcal{N}=1$
quiver.

\section{Global properties and S-duality}
\label{secrev1}

In a general QFT fixing a connected gauge group $\mathbf{G}$ from  gauge Lie algebra $\mathbf{g}$ requires 
to specify additional data, because there are different possible gauge groups corresponding to the same algebra. 
The gauge group is fixed as follows:  consider the universal covering group $\widetilde{\mathbf{G}}$ 
and mod it  by its center $\Gamma$ or a subgroup $\mathbf{H} \subset \Gamma$. 
The gauge group is $\mathbf{G}=\widetilde {\mathbf{G}}/\mathbf{H}$.
This group is commonly referred to as the electric gauge group. Indeed, by keeping in mind the electric-magnetic duality
for Abelian theories, fixing the gauge theory requires also a similar discussion on the magnetic side.
This corresponds to fixing the global properties of the GNO--dual \cite{Goddard:1976qe}
$\mathbf{g}^*$ algebra.
The magnetic group $\mathbf{G}^*$ is not completely fixed yet. Indeed 
for a given group $\widetilde{\mathbf{G}}/\mathbf{H}$, there can be different possible choices of the GNO--dual 
group $\mathbf{G}^*$.

A possible way to fix the gauge group consists of choosing a (maximal) set of allowed line operators.
These line operators are Wilson (W) and 't Hooft (H) lines. 
The W--lines are related to the electric aspects of the gauge theory
while the H--lines are related to the magnetic ones.

Here we consider $\mathcal{N}=4$ SYM with  $\mathbf{g}=su(N)$
and review how line operators are used to specify the global properties of the gauge group.
Once we fix the group  $\mathbf{G}=\widetilde{\mathbf{G}}/\mathbf{H}$
the line operators have to be invariant under $\mathbf{H}$.
Here W--lines are invariant under $\mathbf{H}$, \emph{i.e.} they are labeled by representations
of $\mathbf{G}$. Analogously the H--lines are labeled by representations of $\mathbf{G}^*$.
The spectrum of the allowed operators can be put in 1-1 correspondence with the 
gauge group.
However it is not necessary to specify all the possible representations
but one can restrict to a subset of them, identified by their charge under the center $\Gamma$.
In the rest of this section we review the relations between the charges of the line operators
and the global properties of a gauge group.

When the gauge group is the universal cover $\widetilde{\mathbf{G}}$ there is no restriction on the allowed representations of the Wilson lines.
We can associate the representations to a lattice, the weight lattice $\Lambda_w$ of $\mathbf{g}$, modulo the Weyl group $W$.
Modding  $\widetilde{\mathbf{G}}$ by $\Gamma$ only the representations invariant under $\Gamma$ survive.
They are obtained by moving with the roots of $\mathbf{g}$ on a sub-lattice specified by the adjoint representation. 
The weights of the adjoint are the roots of the algebra $\mathbf{g}$ and this sub-lattice is 
denoted as the root lattice $\Lambda_r$.
One can choose also a subgroup $\mathbf{H} \subset \Gamma$. In this case the representation lives in the 
co-character lattice $\Gamma_{\mathbf{G}^*}$.
The three lattices are related as $\Lambda_r \subset \Gamma_{\mathbf{G}^*} \subset \Lambda_w$.
The center $\mathbf{H}$ and the fundamental group $\pi_1(\mathbf{G})$ are
\begin{eqnarray}
\mathbf{H} = \Gamma_{\mathbf{G}^*}/\Lambda_r,
\quad
\pi_1(\mathbf{G}) = \Lambda_{w}/\Gamma_{\mathbf{G}^*}.
\end{eqnarray}

A similar discussion can be replied for the  GNO--dual algebra $\mathbf{g}^*$. 
The center is $\Gamma^*= \Gamma$
and the Weyl group is $W^*=W$.
The magnetic weight lattice $\Lambda_{mw}$ corresponds to the dual of the root lattice of $\mathbf{g}$.
 The magnetic root lattice is called $\Lambda_{cr}$, and it is usually called the co-root lattice.
 The dual of the co-character lattice is called the  character lattice $
 \Gamma_{\mathbf{G}^*}$. The inclusion here is
 $\Lambda_{cr} \subset \Gamma_{\mathbf{G}^*} \subset \Lambda_{mw}$.
The center $\mathbf{H}$ and the fundamental group $\pi_1(\mathbf{G}^*)$ are
\begin{eqnarray}
\mathbf{H}^* =  \Lambda_{mw}/\Gamma_{\mathbf{G}^*},
\quad
\pi_1(\mathbf{G}^*) = \Gamma_{\mathbf{G}^*}/ \Lambda_{cr}.
\end{eqnarray}
In the case the magnetic set of lines are the 't Hooft lines.
Fixing the gauge group requires to specify the allowed representations of the H--lines
as well.

Observe that there are not only purely electric W--lines or purely magnetic H--lines,
but we can allow also for dyonic (W,H)--lines. 
It turns out that specifying the set of allowed W--lines and H--lines corresponds to specify 
a sub--lattice  $ \Lambda_w \times \Lambda_{mw}$ modulo the action of the Weyl group.

Consider a generic dyonic line $(\lambda_e,\lambda_m) \in  \Lambda_w \times \Lambda_{mw}$,
it is identified to the line $(w \lambda_e,w \lambda_m)$, where $w$ is an element of $W$.
If the line  $(\lambda_e,\lambda_m)$ is allowed also the line  $(-\lambda_e,-\lambda_m)$ is allowed.
Moreover if two lines $(\lambda_e,\lambda_m)$  and $(\lambda'_e,\lambda'_m)$ are allowed 
also the line $(\lambda_e+\lambda'_e,\lambda_m+\lambda'_m)$ is  allowed.
For each choice of $\mathbf{G}$ there is always a pure electric line $(r_e,0)$ with $r_e \in \Lambda_r$
and a pure magnetic line $(0,r_m)$ with $r_m \in \Lambda_{cr}$.

At this level of the discussion we are still considering the electric and magnetic sub-lattices of $\Lambda_w$ and $\Lambda_{mw}$. We can project them to smaller sub-lattices using the following observation.
We can consider a dyonic line $(\lambda_e,\lambda_m)$ in the lattice which specifies a set of points. 
Then we can always add the line $(p \, r_e, q \, r_m)$ with $p,q \in \mathbb{Z}$. In this way we reach all the points 
of the lattice with the same charge as $(\lambda_e,\lambda_m)$ mod the dimension of the center.
This explains why the lines can be organized in classes distinguished by their charge under the center.
A generic point is of the form $(z_e,z_m) \in \Gamma \times \Gamma$.
A theory is specified by a complete set of allowed charges. The lines are obtained from a condition of mutual
locality, corresponding to a DSZ quantization condition on the set of charges.
In the case of $\mathbf{g}=su(N)$,  one has that $\Gamma=\mathbb{Z}_n$ and $\mathbf{H} = 
\mathbb{Z}_{k'}$
where $k' k =N$ and $k,k',N \in \mathbb{N}$.
We can choose the representative of the equivalence class of charges 
by observing that the  fundamental representation has unitary charge under the center. 
A generic point with charge $(e,m)$ can be associated to the representation
$(Sym^e \framebox[2\width]{~} \,;Sym^m \framebox[2\width]{~})$,
\emph{i.e.} the symmetric product of $e$ and $m$ fundamentals. 

The DSZ quantization condition for two pairs of charges $(e_1,m_1)$ and $(e_2,m_2)$ is
\begin{equation}
\label{DSZSU}
e_1 m_2 - m_1 e_2 = 0 \, \,mod \, \, N.
\end{equation}
The gauge group is in general $SU(N)/\mathbb{Z}_k$.
The electric lattice admits only charges $(e_\ell,0)$ with $e_\ell = \ell k $
and $\ell \in \mathbb{Z}$.
There is always the possibility to choose the charge $(0,m_\ell)$ 
with $m_\ell= \ell k'$ and $k'$ integer. This is because $\pi_1(\mathbf{G}) = \mathbb{Z}_{k'}$.
Other choices are anyway possible. They come from the fact that the lattices are generated by the 
DSZ quantization condition. In general the magnetic charge with the smallest $m$ is $(e=\ell,m_{min}=k')$.
There are $k$ different (inequivalent) choices, $\ell=0,\dots,k-1$.
Once this second point in the lattice of the charges is specified we can use the two vectors $(e_{min}=k,0)$ and $(\ell,m_{\min})$
to reconstruct the whole lattice. The reason is that we can combine linearly the lines and then their charges:
if the lines with charge $(e_1,m_1)$ and $(e_2,m_2)$ are in the lattice  then also  the line with charges
$(e_1+e_2,m_1+m_2)$ is in the lattice.
The group associated to the lattice generated by these two vectors is
denoted as $(SU(N)/\mathbb{Z}_k)_\ell$, $\ell=0,\dots,k-1$.
Theories identified by different lattices have different global properties but they are connected by the action of the generators S and T of the $SL(2,\mathbb{Z})$ S-duality group. 
However if $N$ is not square free there are theories that
sit in different orbits of the S-duality group. i.e. self dual theories that cannot be reached by applying any $SL(2,\mathbb{Z})$ transformations.

\section{Lens space index}
\label{review2}

The Lens space index is a generalization of he superconformal index
of \cite{Kinney:2005ej,Romelsberger:2005eg}, computed on
$L(r,1) \times S^1$. The Lens space $L(r,1)$ is an orbifold 
of $S^3$ defined by the identification
\begin{equation}
(z_1,z_2) \sim (z_1e^{\frac{2 \pi i}{r}},z_2 e^{-\frac{2 \pi i}{r}})
\end{equation}
where $z_i$ are complex and the three sphere is defined by
the relation $|z_1|^2 + |z_2|^2=1$.
The index of a chiral multiplet $\Phi$ of $R$-charge $\Delta$ is
\begin{eqnarray}
I_{\Phi}^{(\Delta)}(z)
&=&
\Gamma_e ( (pq)^{\frac{\Delta}{2}} z;q^r,pq)
\Gamma_e ( (pq)^{\frac{\Delta}{2}} p^r z;p^r,pq)
\nonumber\\ 
&=&
\Gamma_e ( (pq)^{\frac{\Delta}{2}} q^r z;q^r,pq)
\Gamma_e ( (pq)^{\frac{\Delta}{2}} z;p^r,pq)
\, ,
\end{eqnarray}
where $\Gamma_e$ represents an elliptic Gamma function 
(see \cite{Dolan:2008qi,Spiridonov:2009za} for a physical definition) 
corresponding to 
\begin{equation}
\Gamma_e(z;p,q) \equiv \prod_{i,j=0}^{\infty} 
\frac{1-p^{i+1} q^{j+1}/z}{1-p^i q^j z}
\, .
\end{equation}
The fugacities $p$ and $q$ refer to the two $SU(2)$ of the original 
$S^3$ while $z$ is a collective fugacity for the other internal symmetries.
In presence of global internal symmetries the index can be refined.
For each global symmetry one can consider a fugacity $z$
and also turn on a non trivial holonomy around a non-contractible cycle
in $L(r,1)$. Going around such cycle the field acquires a phase
$e^{\frac{2 \pi i m}{r}}$.
The net result for the index of $\Phi$ in presence of such an holonomy is
\begin{equation}
I_{\Phi}^{(\Delta)}(m,z) = I_0^{(\Delta)}(m,z) 
\Gamma_e((pq)^\frac{\Delta}{2} q^{r-m} z;q^r,pq)
\Gamma_e((pq)^\frac{\Delta}{2} p^{m} z;p^r,pq) \, ,
\end{equation}
where $I_0$ represents the vacuum energy, that, for non-anomalous symmetries (the only one that will be considered here) is  
\begin{equation}
I_0^{(\Delta)}(m,z) = 
\left(\frac{(pq)^{1-\Delta}}{z^2}\right)^\frac{m(r-m)}{4r}
\left(\frac{p}{q}\right)^\frac{m(r-m)(r-2m)}{12r}
.
\end{equation}
The contribution of an $\mathcal{N}=1$ vector multiplet to the Lens index is
\begin{equation}
I_V(m,z) = \frac{I_0^V(m,z)}{(1-1/z)^{\delta_{m,0}}
\Gamma_e(q^m/z;q^r,pq) \Gamma_e(p^{r-m}/z;p^r,pq)} \, ,
\end{equation}
where 
\begin{equation}
I_0^V(m,z) =  \left(\frac{pq}{z^2}\right)^{-\frac{m(r-m)}{4r}}
\left(\frac{q}{p}\right)^\frac{m(r-m)(r-2m)}{12r}
\, .
\end{equation}

If there is also a gauge group one has to properly sum
the contribution of the non-trivial holonomies on the matrix
integral. 
The path integral over the gauge group $\mathbf{G}$ localizes indeed onto flat connections, labeled by the holonomies around the $S^1$ and the cycle inside $L(r,1)$. These holonomies are elements of the gauge group and can be labeled by $h$ and $g$,
provided that they 
respect the conditions $g^r=1$ and $g h g^{-1} h^{-1} =1$.
The holonomies must be considered up to simultaneous conjugation, leading to discrete set of choices for $g$. One can fix $g$ to lie in the maximal torus.
This is the end of the story for simply connected groups.
In this case one obtain the Lens space index by summing over the inequivalent discrete choices of $g$, by integrating over the eigenvalues of $h$ and by properly considering the Haar measure of the gauge group preserved by $g$ in the matrix integral.

Let us consider the case of $\widetilde{\mathbf{G}} = SU(N)$, even if the situation is more general.
If one considers a non simply connected gauge group $\mathbf{G}=SU(N)/\mathbb{Z}_d$, obtained by modding the universal covering group $\widetilde{\mathbf{G}}=SU(N)$ by a (sub)group $\mathbf{H}=\mathbb{Z}_d$ 
of the center $\Gamma=\mathbb{Z}_N$, the situation  is more complicated.
The holonomies in this case have to satisfy the following constraints
\begin{equation}
\label{commutation}
g^r = \mu = e^\frac{2 \pi i k}{N}\quad \text{and}  \quad g h g^{-1} h^{-1} = \nu
=  e^\frac{2 \pi i \ell}{N} \, ,
\end{equation}
where $k$ and $\ell$ are integer and $k \simeq k+\text{gcd}(N,r)$ and $\ell = \hat \ell N/\text{gcd}(N,r)$, such that $k$ and $\hat \ell$ 
can be chosen to lie between $0$ and $\text{gcd}(N,r)-1$.
In the rest of the discussion we fix $r=N$, that is the simplest
case in which the global properties give different contributions to the integrand.
These holonomies are referred as almost commuting
 and their contribution has to be computed separately. In such cases the integral over the 
gauge group is reduced, because, depending on the choices
of $\mu$ and $\nu$, some of the eigenvalues of $h$ are
fixed. 
In the index the contribution of the fields is usually
obtained by decomposing them in the weight space.
If $g$ and $h$ are almost commuting they do not lie in the maximal torus, and it implies that the action on the weight basis is not necessarily diagonal. In this case one has to find
a basis that makes such an action diagonal and sum over the eigenvalues in this basis.
In the next section we will review the choice of the basis for
the case of the adjoint representation of $SU(N)$.
The index becomes a sum over the sectors identified by different values of $\mu$ and $\nu$.
Each sector is weighted by a discrete theta angle,
identified by a phase involving rational number $c(w_2)$
where $w_2$ represents  the second Stiefel-Whitney class.
The index is formally
\begin{equation}
\label{formal}
I = \frac{1}{|\mathbf{H}|} \sum_{\mu,\nu} e^{i c(w_2(\mu,\nu))} Z_{\mu,\nu} \, .
\end{equation}

\subsection{$\mathcal{N}=4$ SYM and global properties}

Here we review the results of \cite{Razamat:2013opa} for the derivation of the Lens space index for the case of  $SU(N)$ $\mathcal{N}=4$ SYM. This is necessary to set the notations  for the derivation of the index in cases with lower supersymmetry.

Let us first review the results for the sector $\hat \ell=0$ with a generic $k$.
In this case the matrices $\tilde g$ and $\tilde h$ are both diagonal. The elements of $\widetilde{g}$ are
$e^{2 \pi i/r(m_i+k/N)}$ with $m_1+\dots +m_N+k=0$ mod $r$ and one can fix
$r>m_1\geq m_2 \dots \geq m_N \geq 0$.
Each $\mathcal{N}=1$ adjoint chiral multiplet $\Phi_i$ contributes as
\begin{equation}
I_{\Phi}^{(\Delta)} = I_\Phi^{(\Delta)} (0,1)^{N-1} \prod_{i \neq j} I_\Phi^{(\Delta)} ([m_i-m_j],z_i z_j^{-1})
\end{equation}
where 
$[m]\equiv n \, | 0 \leq n \leq r-1, \, n=m$ mod $r$.
I the following we omit to indicate $\Delta$, because in models with non-abelian R-symmetry 
we always have $\Delta=2/3$.
The $\mathcal{N}=1$ vector multiplet contributes as
 \begin{equation}
I_{V} = I_V (0,1)^{N-1} \prod_{i \neq j} I_V ([m_i-m_j],z_i z_j^{-1})
\end{equation}
and the measure factor is
\begin{equation}
\Delta_{\vec m}(z) = \frac{1}{|W_{\vec m,N}|} \prod_{i \neq j}(1-z_i/z_j)^{\delta_{[m_i-m_j],0}} 
\end{equation}
where $|W_{\vec m,N}|$ represents the size of the Weyl group left unbroken.

The next step consist of computing the contribution from the almost commuting holonomies, coming from the sectors with $\hat \ell \neq 0$, i.e. $\nu \neq 1$.
This was done in \cite{Razamat:2013opa} as follows. First one studies the case where $N$ and $\ell$ are coprime.
In this case the almost commuting holonomies can be represented by a pair of matrices denoted
as  "clock"  and "shift" matrices, the first one corresponding to $C_{\ell,N} = 
\epsilon_N^\ell$diag$(e^\frac{2  \pi i \ell j }{N})$
with $j=0,\dots, N-1$ and the second one $S_{N} =  \epsilon_N \delta_{i+1,i}$ (where $\epsilon_N$ is necessary for $\tilde g$ and $\tilde h$ to have unit determinant). 
These are the unique matrices, up to simultaneous conjugation and
multiplication by a scalar, satisfying (\ref{commutation}).

By identifying the clock and the shift matrices with the matrices $\tilde g$ and $\tilde h$, 
the contribution of from such holonomies can be computed.
The computation requires a change of basis for the matter and vector 
fields in the adjoint representation, because the action of $\tilde g$ and $\tilde h$ is not diagonal
in the weight basis.
By denoting the weight basis as $E_{i,j} = e_i \otimes e_j^*$, where $e_i$ represents base element of $\mathbb{C}^N$,
the new basis is 
\begin{equation}
 F_{m,n} = \sum_j e^{2 \pi i m j/N} E_{j,n-j} \, .
 \end{equation}
The traceless condition $\sum_i E_{i,i} = 0$ becomes $F_{0,0}=0$ in this new basis.
The adjoint action of $\tilde g$ and $\tilde h$  is diagonal in this basis, with eigenvalues given by 
\begin{equation}
\tilde g \cdot F_{m,n}  = e^{2 \pi i n \ell /N}  F_{m,n} 
\quad \text{and} \quad
\tilde h  \cdot F_{m,n}  = e^{2 \pi i m /N}  F_{m,n} \, .
\end{equation}
The index for an $\mathcal{N}=1$ adjoint chiral fields is obtained by considering the allowed values 
of $m$ and $n$.
The contribution to the index of a vector multiplet and of an adjoint chiral field are
\begin{eqnarray}
{\prod_{m,n=0}^{N-1}}'  I_{V}^{} \left(\frac{n \ell r}{N}, e^{2 \pi i m/N} \right)
\quad
\text{and}
\quad
{\prod_{m,n=0}^{N-1}}'   I_{\phi}^{} \left(\frac{n \ell r}{N}, e^{2 \pi i m/N} \right)
\end{eqnarray}
respectively,
where the apex denotes the traceless condition, \emph{i.e.} we have to avoid to consider $m$ and $n$  
vanishing simultaneously.
In this case there is no integral to perform over the gauge group, because the matrix $\tilde h$ has been completely fixed.

In general one has to consider cases in which $N$ and $\ell$ are not coprime. Let us define $d=\text{gcd}(N,\ell)$.
In this case the holonomies can be written in the form
\begin{equation}
\tilde g = \text{diag} \{ e^{\frac{2 \pi i}{r}(m_i+\tilde k/N)} \} \otimes C_{\ell/d,N/d}
\quad \text{and} \quad
\tilde h = \text{diag}  \{ z_i\} \otimes S_{N/d}
\end{equation}
where 
$\tilde k = k$ for $N/d$ odd and $\tilde k = k- rd/2$ for $N/d$ even.
The only pairs allowed satisfy the condition $\tilde k \ell = 0$ mod $N$.
In this case one can turn to a diagonal basis by a tensor product of the weight basis and 
of the $F_{m,n}$ basis. This gives a basis $F_{m,n;i,j}$ with $m,n = 0,\dots N/d-1$ and
$i,j=1,\dots,d$.
The eigenvalues are
\begin{equation}
\tilde g \cdot F_{m,n;i,j} = e^{2 \pi i n \ell /N+(m_i-m_j)/r}  F_{m,n;i,j} 
\quad \text{and} \quad
\tilde h \cdot F_{m,n;i,j}  = e^{2 \pi i m /N} z_i/z_j  F_{m,n;i,j} 
\end{equation}
and for each allowed value of $m,n,i$ and $j$ we have a contribution to the index of the form
$ I_\Phi([n \ell r/N+m_i-m_j],e^{2\pi i m d/N} z_i/z_j) $
where $[m] \equiv n | 0 \leq m \leq r-1$, $n=m$ mod $r$.
Analogous contributions can be found from the vector multiplet
and for the measure factor.

Eventually one can collect all the results reviewed above and write the general form of the index.
The final form of the index given in in (\ref{formal})
where the different terms  $Z_{\mu,\nu}$  get multiplied by a phase in the partition function.
This phase depends on the discrete theta angle and in this case, as explained in \cite{Razamat:2013opa}, 
we have
\begin{equation}
\label{LSI}
I = \frac{1}{d} \sum_{\hat k, \hat \ell=0}^{d-1} e^{2 \pi i n \hat k \hat \ell/d} Z_{\tilde k,  \ell}
\end{equation}
where $\tilde k = \hat k N/d$ and $ \ell= \hat \ell N/d$.
Each sector contributes as
\begin{eqnarray}
Z_{\tilde k,\ell} &=&
\sum_{m_i} \,
\oint \prod_{i=1}^{N}
\frac{d z_i}{2 \pi i z_i}  
\Delta_{\vec{m}}^\ell(z) \quad
\\
&\times&
{\prod_{m,n=0}^{N/d-1}}'
\prod_{i,j=1}^d
 \left( I_{V}\big(\big[\frac{n \ell r}{N}+m_{i,j}\big], e^{\frac{2 \pi i m d}{N}} z_{i,j}\big)
 \prod_{I=1}^{3} I_{\Phi_{I}}^{(\frac{2}{3})}\big(\big[\frac{n \ell r}{N}+m_{i,j}\big], e^{\frac{2 \pi i m d}{N}}  z_{i,j}\big)\right)
\nonumber 
\end{eqnarray}
where we defined $z_{i,j} \equiv z_i/z_j$ and $m_{i,j} \equiv m_i-m_j$
and the sums over $m_i$ are restricted by $\sum_i m_i = - \tilde k d / N (\,\text{mod}\, r d /N)$.
The notation  $\oint$ refers to the constraint $\prod_{i=1}^{N} z_i =1$.
Observe that this index can be refined by adding two further fugacities that distinguish the different 
chiral adjoint fields, but here we omitted this contribution because such a refinement  it is not
necessary for our discussion.

\section{The Lens  space index for $\mathcal{N}=2$ elliptic models}
\label{sec:n2LSI}

In this section we study the Lens space index for infinite families of
quiver gauge theories arising from M-theory or type IIA.
In M-theory these models describe a stack of $N$ M5 branes wrapped
on a genus $g=1$ Riemann surface with $n_G$ punctures.
In type IIA the model is related to a stack of $N$ D4 branes extended along
the directions $01236$, where the direction $x_6$ is compact.
There are also  $n_G$ parallel NS5 branes 
placed at different positions $x_6^i$ along the circle.
On the field theory side the model corresponds to a necklace $\mathcal{N}=2$ 
quiver gauge theory with $n_G$ gauge groups $SU(N)$ and one bifundamental
hypermultiplet connecting each pair of consecutive nodes.

It was realized in \cite{Amariti:2016hlj} that the gauge group of the theory is actually determined once
additional data on the (W,H) lines are specified.
This is because the original gauge group is $U(N)_1 \times U(N)_2  \times \dots \times U(N)_{n_G}$
and in the IR $n_G-1$ combinations of the $U(1)$'s
become baryonic and the remaining overall $U(1)$ is decoupled.
The inequivalent ways in which such an $U(1)$ decouples give raise to different
gauge groups identified by their global properties.
By borrowing the notations adopted for $\mathcal{N}=4$ SYM
the gauge group here is specified by
\begin{equation}
\label{additional}
G = \bigg( \bigg ( \prod_{j=1}^{n_G} SU(N)_j\bigg)/\mathbb{Z}_k   \bigg)_i
.
\end{equation}
The bifundamentals break the center $(\mathbb{Z}_N)^{n_G}$ to its diagonal
subgroup $\mathbb{Z}_N^{\text{diag}}$,  implying $\mathbb{Z}_k \subset \mathbb{Z}_N^{\text{diag}} $ in  (\ref{additional}).
One can connect theories with different global properties
by the action of the $SL(2,\mathbb{Z})$ group, that represents a subgroup of the full
S-duality group, identified with the mapping class group of the $n_G$-punctured Riemann 
surface \cite{Halmagyi:2004ju}. 
Analogously to the case of $\mathcal{N}=4$ SYM, if $N$ is not square free there are orbits 
in the action of this S-duality subgroup.
In the following we show how to compute the Lens space index for this family of theories,
matching  the index under S-duality for some low rank case
and verifying the existence of the S-duality orbits. 

We consider, as in the case of $\mathcal{N}=4$ SYM,  only the $\mathcal{N}=1$ index of \cite{Benini:2011nc}. 
The corresponding 
$\mathcal{N}=1$ $R$-current is generated by the combination of the abelian 
generators of the $R$-symmetry and it assigns $R$-charge $2/3$ to each matter field.
Each $su(N)$ factor has center $\Gamma=\mathbb{Z}_N$, but in this case
the matter fields break it to its diagonal subgroup. It modifies also the holonomies
$g$ and $h$ contributing to the index.

In general we can re-organize the matrices $g$ and $h$ as $n_G N \times n_G N$ 
block matrices, each with $n_G$ diagonal block of dimension
$N \times N$
\begin{equation}
g = \left(
\begin{array}{cccc}
g_1 &0 &0 &0 \\
0 & \dots &0&0\\
0 &0&g_{n_G-1}& 0\\
0&0&0&g_{n_G}
\end{array}
\right)
\quad
\quad
h = \left(
\begin{array}{cccc}
h_1 &0 &0 &0 \\
0 & \dots &0&0\\
0 &0&h_{n_G-1}& 0\\
0&0&0&h_{n_G}
\end{array}
\right)\, .
\end{equation}
The constraints (\ref{commutation}) must be satisfied by the holonomies 
in each of these $n_G$ sectors separately.
At each block we have the relations
\begin{equation}
\label{commutation2}
g_I^r = \mu = e^\frac{2 \pi i k}{N}, \quad g_I h_I g_I^{-1} h_I^{-1} = \nu
=  e^\frac{2 \pi i \ell}{N} 
\end{equation}
for $I=1,\dots,n_G$.
A crucial observation is that  the $2n_G$ matrix equations (\ref{commutation2})
have to be satisfied  for the same value of $k$ and $\ell$ .
This is because only the diagonal subgroup of the center survives in presence
of the bifundamental hypermultiplets.
This allows us to organize the computation as above, 
by separating the contribution of the sectors
with the integers $k$ and $\ell$.

Let us start by discussing the index for the $Z_{k,\ell}$ sector where $\ell=0$. 
For each $k$ the index is 
\begin{eqnarray}
Z_{k,0} &=& \sum_{m_i^I} 
(I_{V_I}(0,1)\,
I_{\Phi_I}^{}(0,1))^{(N-1)n_G} 
\oint 
\prod_{I=1}^{{n_G} } 
\prod_{i=1}^{N} 
\frac{d z_i^I}{2 \pi i z_i^I}  
\prod_{I \rightarrow J}
I_{\Phi_{I,J}}^{}([m_{i,j}^{I,J} ],z_{i,j}^{I,J})
\nonumber \\
&\times&
\prod_{I=1}^{n_G} 
\prod_{1\leq i<j \leq N} 
\bigg(\frac{1}{2} (1-z_{i,j}^{I,I})^{\pm 1}
 I_{V_I}\left([m_{i,j}^{I,I}], (z_{i,j}^{I,I})^{\pm 1}\right)
 I_{\Phi_I}^{}\left([m_{i,j}^{I,I} ], (z_{i,j}^{I,I})^{\pm 1}\right) \!\!\!
 \bigg)
\end{eqnarray}
where we defined $m_{i,j}^{I,J} \equiv m_i^I-m_j^J$ and $z_{i,j}^{I,J} \equiv z_i^I/z_j^J$.
Furthermore the notation $I \rightarrow J$  refers to the bifundamentals 
connecting the nodes of the quiver 
(for example here the only possibilities are $J = I +1$ and $J=I-1$).
We also used the notation $f(x^{\pm 1}) \equiv f(x) f(x^{-1})$.
In addition we are not refining the index, i.e. all the chemical potentials
for the global symmetries are set to zero.

We can now choose each $g_I$-block as done by \cite{Razamat:2013opa}, by constraining the
sums over $\vec{m}^I$ . Here we must consider each $\vec{m}^I$ such that $\sum_i m_{i}^{I} + k = 0 \text{ mod }r$,
where $i$ runs from $1$ to $N$ and $I=1,\dots,n_G$.
The choice $\ell=0$ fixes $h_I$ = \text{diag}$\{z_i^I\}$,
where $z_i^I$ are defined on the circle and $\prod_i z_i^I=1$ for each $I$.

The sector with $\ell \neq 0$ has to be analyzed by diagonalizing the holonomies
in a non conventional basis.
Let us consider separately the contribution of the adjoints and of the bifundamentals.
First we analyze the case in which $d=1$, where $d=\text{gcd}(N,\ell)$.
In this case the matrices $g$ and $h$ have again two diagonal blocks
of the type discussed in \cite{Razamat:2013opa}.

The novelty in this case is the presence of the bifundamentals.
They are defined by the representations
$E^{(I,J)}_{i,j} = e_i^I \otimes  {e_j^J}^*$ in the weight basis.
A diagonal basis in this case is
\begin{equation}
F_{m,n}^{(I,J)} = \sum_j e^{2 \pi i m j/N} E_{j,n-j}^{(I,J)}\, .
\end{equation}
The eigenvalues are
\begin{equation}
g \cdot F_{m,n}^{(I,J)}  = e^{2 \pi i n \ell /N}  F_{m,n}^{(I,J)} 
\quad \text{and}
\quad
h \cdot F_{m,n}^{(I,J)}  = e^{2 \pi i m /N}  F_{m,n}^{(I,J)} \,.
\end{equation}
The contribution to the index is 
\begin{equation}
{\prod_{m,n=0}^{N-1}}'  I_{\phi_{I,I}} \left(\frac{n \ell r}{N}, e^{2 \pi i m/N} \right)
\end{equation}
for the adjoints and 
\begin{equation}
\prod_{m,n=0}^{N-1}  I_{\phi_{I,J}} \left(\frac{n \ell r}{N}, e^{2 \pi i m/N} \right)
\end{equation}
for the bifundamentals, where in the second case we did not exclude the 
value $n=m=0$, because there is no traceless condition on these fields.
The contribution of the vector multiplets can be computed straightforwardly in this basis, following 
\cite{Razamat:2013opa}.

The only step that is missing is the calculation of the contributions of the sectors with $d > 1$.
Also in this case one can compute them following the strategy above and applying it to the 
results of \cite{Razamat:2013opa}.

The calculation is identical until formula (5.26) of \cite{Razamat:2013opa}:
 in this case the basis is $F_{m,n;i,j}^{(I,J)}$
and equation (5.26) is modified as
\begin{equation}
g \cdot F_{m,n;i,j}^{(I,J)}  = e^{2 \pi i n \ell /N + (m_i^I - m_j^J)}  F_{m,n;i,j}^{(I,J)} 
\quad
\text{and} 
\quad
h \cdot F_{m,n;i,j}^{(I,J)}  = e^{2 \pi i m d/N} z_i^I/z_j^J  F_{m,n;i,j}^{(I,J)} 
\end{equation}
and the traces correspond to $\sum_i F_{0,0;i,i}^{(I,I)}$.
Again the contribution of the adjoint requires to not consider $m=n=0$ 
while for the bifundamentals such term contributes.
By summing the various contribution togheter 
the generic expression for $Z_{k,\ell}$ of an elliptic model
with $n_G$ gauge groups is
\begin{eqnarray}
\label{Zkln2}
Z_{\tilde k,\ell} &=&
\sum_{m_i^I}
\oint \prod_{I=1}^{n_G} 
\frac{d z_i^I}{2 \pi i z_i^I}  
\Delta_{\vec{m}^I}^\ell(z^I)
{\prod_{m,n=0}^{N/d-1}} \prod_{i,j=1}^d
\prod_{I \rightarrow J}
I_{\Phi_{I,J}}\big(\big[\frac{n \ell r}{N}+m_{i,j}^{I,J} \big], e^{\frac{2 \pi i m d}{N}} 
z_{i,j}^{I,J}\big)
\nonumber  \\
&\times&
{\prod_{m,n=0}^{N/d-1}}' \prod_{i,j=1}^d
\left(
 I_{V_I}\big(\big[\frac{n \ell r}{N}+m_{i,j}^{I,I}\big], e^{\frac{2 \pi i m d}{N}} z_{i,j}^{I,I}\big)
 I_{\Phi_{I}}\big(\big[\frac{n \ell r}{N}+m_{i,j}^{I,I}\big], e^{\frac{2 \pi i m d}{N}}  z_{i,j}^{I,I}\big)
 \right)
\nonumber \\
\end{eqnarray}
where the sum over $m_i^I$ is constrained by 
$ \sum_i m_i^I = - \tilde k d / N (\text{ mod } r d /N)$.
The index is obtained by applying (\ref{LSI}).
Again we did not consider the presence of flavor fugacities, but they can be turned on 
straightforwardly.
\\
\\

In the following we  focus on the example with $n_G=2$. Despite the simplicity of this model
its Lens space index contains all the necessary ingredients 
of the one expected for generic $n_G$. 
In this case we have two gauge groups $SU(N)_1 \times SU(N)_2$
and two bifundamental hypermultiplets. In $\mathcal{N}=1$ notations
there are two adjoint matter fields $\Phi_1$ and $\Phi_2$, two bifundamentals
$a_1$ and $a_2$ and two antifundamentals $b_1$ and $b_2$. 
The superpotential is 
\begin{equation}
W = a_1 \Phi_1 b_1 - a_2 \Phi_1 b_2 - a_1 \Phi_2 b_1 + a_2 \Phi_2 b_2 \, .
\end{equation}
By turning off the global symmetries the model has an $SU(2) \times U(1)$
$R$-symmetry.
Here we only consider the $\mathcal{N}=1$ $R$-symmetry 
that assigns charges $R=2/3$ to the adjoints and the bifundamental fields.
It corresponds to $R_{\mathcal{N}=1} = \frac{4}{3} J_3+\frac{1}{3} R_{\mathcal{N}=2}$,
where $J_3$ is the diagonal generator of $SU(2)$.
We provide the explicit calculation of the Lens space index 
$I_{L(r,1) \times S^1}$ for this  model.
We focus on the case $r=N$ and explicitly compute the cases $N=2,3,4$.
In the first two case, $N=2$ and $N=3$ we found that the index coincides,
signaling that the models are related by S-duality. In the last case we find that
the Lens space index differs in the two orbits of the S-duality groups.

\subsection*{The $SU(2) \times SU(2)$ case}

\begin{equation}
\begin{array}{cc}
\begin{array}{|c|c|c|}
\hline
\ell  \backslash \tilde k&0&1\\
\hline
0 &a&b\\
\hline
1 &c&0\\
\hline
\end{array}
&\hspace{.2cm}
\begin{array}{ccccccccccccccc ccc}
a &=& 2&+&26 x^{4/3}&-&12 x^2&+&98 x^{8/3}&-&100 x^{10/3}&+&246 x^4 &+&\dots\\
b &=& 1&+&2 x^{2/3}&+&17 x^{4/3}&+&4 x^2&+&66 x^{8/3}&-&70 x^{10/3}&+&234 x^4
 &+&\dots\\
c &=& 1&-&2 x^{2/3}&+&9 x^{4/3}&-&16 x^2&+&32 x^{8/3}&-&30 x^{10/3}&+&12 x^4
 &+&\dots\\
\end{array}
\end{array}
\end{equation}
\begin{equation}
\begin{array}{l}
I_{SU(2) \times SU(2)}=a=2+26 x^{4/3}-12 x^2+98 x^{8/3}-100 x^{10/3}+246 x^4
\\
I_{((SU(2) \times SU(2))/\mathbb{Z}_2)_{0,1}} 
=\frac{a+b+c}{2} =2+26 x^{4/3}-12 x^2+98 x^{8/3}-100 x^{10/3}+246 x^4+
\dots
\end{array}
\end{equation}
As expected the index for the $SU(2) \times SU(2)$ case coincides with the one 
of the $(SU(2) \times SU(2))/\mathbb{Z}_2)_i$ cases, for both $i=0$ and $i=1$.

\subsection*{The $SU(3) \times SU(3)$ case}

\begin{equation}
\begin{array}{cc}
\begin{array}{|c|c|c|c|c|}
\hline
\ell  \backslash \tilde k&0&1&2\\
\hline
0 &a&b&b\\
\hline
1 &c&0&0\\
\hline
2 &c&0&0\\
\hline
\end{array}
&\hspace{1cm}
\begin{array}{ccccccccccccc}
a &=& 4&+&4 x^{2/3}&+&42 x^{4/3}&+&86 x^2&+&183 x^{8/3}&+&\dots  \\
b &=& 3&+&x^{2/3}&+&39 x^{4/3}&+&84 x^2&+&192 x^{8/3}&+&\dots  \\
c &=& 1&-&2 x^{2/3}&+&3 x^{4/3}&+&2 x^2&-&9 x^{8/3}&+&\dots  \\
\end{array}
\end{array}
\end{equation}

\begin{equation}
\begin{array}{l}
I_{SU(3) \times SU(3)}=a=
4+4 x^{2/3}+42 x^{4/3}+86 x^2+183 x^{8/3}+\dots
\\
I_{((SU(3) \times SU(3))/\mathbb{Z}_3)_{0,1,2}} 
=\frac{a+2 b+2 c}{3} =
4+4 x^{2/3}+42 x^{4/3}+86 x^2+183 x^{8/3}+\dots
\end{array}
\end{equation}
Also in this case we observe that the index for the $SU(3) \times SU(3)$ case coincides with the one 
of the $(SU(3) \times SU(3))/\mathbb{Z}_3)_i$ cases, for  $i=0,1,2$.

\subsection*{The $SU(4) \times SU(4)$ case}

\begin{equation}
\begin{array}{cc}
\begin{array}{|c|c|c|c|c|}
\hline
\ell  \backslash \tilde k&0&1&2&3\\
\hline
0 &a&b&c&b\\
\hline
1 &d&0&0&0\\
\hline
2 &e&0&f&0\\
\hline
3 &d&0&0&0\\
\hline
\end{array}
&\hspace{1cm}
\begin{array}{l l c l cl c l c l l}
a  &=&10 &+& 20 x^{2/3} &+& 162 x^{4/3} &+& 320 x^2 &+&\dots  \\
b  &=& 8 &+&  24 x^{2/3} &+&  152 x^{4/3} &+&  336 x^2 &+&\dots \\ 
c  &=& 9 &+& 22 x^{2/3} &+& 159 x^{4/3}  &+& 324 x^2&+&\dots\\
d  &=& 1 &-& 2 x^{2/3} &+& 3 x^{4/3}  &-& 4 x^2  &+&\dots\\
e  &=& 2 &-& 4 x^{2/3} &+& 10 x^{4/3} &-&  16 x^2  &+&\dots\\
f   &=& 1 &-& 2x^{2/3} &+& 7 x^{4/3} &-& 12 x^2&+&\dots
\end{array}
\end{array}
\end{equation}

\begin{equation}
\begin{array}{l}
 I_{SU(4) \times SU(4)} = a=10+20 x^{2/3}+162 x^{4/3}+320 x^2 + \dots  \\
    I_{((SU(4) \times SU(4))/\mathbb{Z}_2)_0} =\frac{a+c+e+f}{2} 
   =11+18 x^{2/3}+169 x^{4/3}+308 x^2+ \dots\\
     I_{((SU(4) \times SU(4))/\mathbb{Z}_2)_1} =\frac{a+c+e-f}{2} =10+ 20 x^{2/3}+162 x^{4/3}+320 x^2 \dots\\
    I_{((SU(4) \times SU(4))/\mathbb{Z}_4)_i} = \frac{a+2 b+c+2 d+e+f}{4} =10+20 x^{2/3}+162 x^{4/3}+320 x^2+\dots
\end{array}
\end{equation}
where in the last line $i=0,\dots,3$.
Here the situation is more interesting because  not all the indices coincide. This is because the
lattice of charges associated to the gauge group 
$\mathbf{G}=((SU(4) \times SU(4))/\mathbb{Z}_2)_0$ is self dual under $SL(2,\mathbb{Z})$.
This model represents an S-duality orbit.
The other orbit includes all the other gauge groups, because their lattices are 
all related by  $SL(2,\mathbb{Z})$ transformations.
Here we have confirmed this expectation showing that the Lens space index 
assumes different values on the two orbits. This provides a non trivial check 
of S-duality in this model.

\section{Conclusions}
\label{conclusions}
In this paper we have studied the Lens space index for a family of $\mathcal{N}=2$ 
elliptic models with unitary gauge groups. We have computed explicitly such an
index for the model with two gauge groups and for low ranks, verifying the action
of the S-duality group.

One could try to extend the analysis to other quiver gauge theories, either 
with real gauge groups or with a reduced amount of
supersymmetry or with tensorial matter.
In each case the global properties are specified by the additional data, and 
the Lens space index should reproduce this structure. 
It may be useful to study the role of S-duality in these cases.
Another extension of this work regards the study of other $\mathcal{N}=2$ 
models as for example class S-theories \cite{Gaiotto:2009we}.
It should be interesting apply the results of
\cite{Razamat:2013jxa,Alday:2013rs,Fluder:2017oxm}, like
the relation between the twisted character of the chiral algebra
and the Lens space index, in order to  reproduce the lattices and the S-duality structure obtained in 
\cite{Drukker:2009tz,Xie:2013lca,Tachikawa:2013hya,Xie:2013vfa,Bullimore:2013xsa,Tachikawa:2015iba,
Coman:2015lna,Amariti:2015dxa,Amariti:2016bxv}.
Also the results of  \cite{Garcia-Etxebarria:2019cnb}, investigating the geometric origin of the 
global properties from the 6d $\mathcal{N}=(1,0)$ perspective, may be useful for the study of 
$\mathcal{N}=2$ models.

\section*{Acknowledgments}

We are grateful to Domenico Orlando and Susanne Reffert for early collaboration
on this project. This work has been supported
in part by Italian Ministero dell'Istruzione, Universit\`a e Ricerca (MIUR),
in part by Istituto
Nazionale di Fisica Nucleare (INFN) through the ``Gauge Theories, Strings,
Supergravity" (GSS) research project and in part by
MIUR-PRIN contract 2017CC72MK-003.

\appendix
\section{$\mathcal{N}=1$ theories}
\label{toriclines}

In this section we discuss the global properties of a family of 4d $\mathcal{N}=1$ 
SCFTs. These theories are quiver gauge theories with bifundamental matter fields
connecting the various nodes. Despite the fact that in presence of matter fields 
the center of the various gauge factors is (partially) broken, we show that
the theories discussed here have the same structure for the global properties of 
$\mathcal{N}=4$ SYM and of the $\mathcal{N}=2$ elliptic models.

Let us first introduce the setup. We consider the gauge theories living on $N$ D3 branes on toric CY-threefolds, corresponding
to  $\mathcal{N}=1$ SCFTs that can be represented by a quiver.
This quiver is a connected graph, in which the nodes represent the gauge groups and 
the oriented arrows are bifundamental fields connecting the $I$-th to the $J$-th node.
A crucial property of such a quiver is that  it can be represented as a planar graph on a torus.
In this case each closed path represent a superpotential term.
Having a connected graph on a torus is enough to argue the existence of pair of
paths connecting each pair on nodes with opposite orientation.
This has a consequence on the line operators.
Let us define  the charge of a generic line as  $(\vec e;\vec m) = (e_1,\dots,e_{n_G};m_1,\dots,m_{n_G})$.
The matter fields impose that $m_I = m_J$ for each pair of integers $I,J$ from $1$ to $n_G$.
On the other hand an electric line ($\vec e;\vec 0)$ can always be put in the form 
$(\sum_{I=1}^{n_G} e_I,0,\dots,0;\vec 0)$,
where $\sum_{I=1}^{n_G} e_I$ must be considered mod $N$.
Again this structure for the charges of the lines allows different lattices, generated by the vectors
$(k,0)$ and $(i,k')$ with $k k' = N$ and $0 \leq i < k$.
This implies that also in the case of $\mathcal{N}=1$ toric quiver gauge theories the gauge group is
\begin{equation}
G = \bigg( \prod_{i=1}^{n_G} SU(N)/\mathbb{Z}_k \bigg)_i\,.
\end{equation}
The  $SL(2,\mathbb{Z})$ symmetry that allows to pass from one to another lattice 
acts on the diagonal holomorphic gauge coupling $\tau_{\text{diag}}$.
It can be identified with the dilaton $\phi$ of the AdS$_5 \times X_5$ string theory dual
background, combined with the theta angle.
This can be reformulated in geometric terms by considering the gauge theory from the 
perspective of the brane tiling. We have  $N$ D5$_{012357}$, one NS5$_{012345}$
and one NS5$_{012367}$. The diagonal holomorphic coupling is identified by the combination 
$e^{C_{57}} + i e^{-\phi}$.
This corresponds to a marginal deformation identified in \cite{Imamura:2007dc}, common to all the theories in this class.
In particular it corresponds to the holomorphic gauge coupling if $X_5 = S^5$, i.e. if we consider 
$\mathcal{N}=4$ SYM.
It is then natural to expect that the $SL(2,\mathbb{Z})$ action allowing to connect the various lattices of the 
$\mathcal{N}=4$ theory acts in a similar manner in the generalization to $\mathcal{N}=1$.
Observe that another reason for such similarity consists of the presence of an overall 
$U(1)$ in the quiver under which every field is uncharged. 
The different ways to decouple this $U(1)$ symmetry correspond to the different 
lattices on the field theory side \cite{Moore:2014gua}.
An interesting consequence of the construction is the fact that also in the case of 
$\mathcal{N}=1$ theories one expect orbits in the action of $SL(2,\mathbb{Z})$ on 
$\tau_{\text{diag}}$.

Here, motivated by this expectation, we study the Lens space index for  these $\mathcal{N}=1$ 
quiver gauge theories.
The derivation of the index follows the one discussed in the body of the paper
and we will not repeat it here.
The only differences are the fact that in this case the quiver is not necessarily vector like
and that the $R$-charges of the bifundamentals are not necessarily $R_{\Phi_{IJ}}=2/3$.
For a generic $\mathcal{N}=1$  quiver each sector specified by $\tilde k$ and $\ell$
contributes to the index as 
\begin{eqnarray}
Z_{\tilde k,\ell} &=&
\sum_{m_i^I }
\oint \prod_{I=1}^{n_G} 
\frac{d z_i^I}{2 \pi i z_i^I}  
\Delta_{\vec{m}^I}^\ell(z^I)
{\prod_{m,n=0}^{N/d-1}}' \prod_{i,j=1}^d
 I_{V_I}\big(\big[\frac{n \ell r}{N}+m_{i,j}^{I,I}\big], e^{\frac{2 \pi i m d}{N}} z_{i,j}^{I,I}\big)
\nonumber
\\
&\times&
\prod_{I \rightarrow J}
I_{\Phi_{I,J}}^{(\Delta_{I,J})}(0,1)^{-\delta_{I,J}}
{\prod_{m,n=0}^{N/d-1}} \prod_{i,j=1}^d
I_{\Phi_{I,J}}^{(\Delta_{I,J})}\big(\big[\frac{n \ell r}{N}+m_{i,j}^{I,J} \big], e^{\frac{2 \pi i m d}{N}} \!\!\! \mu_{\{I,J\}} z_{i,j}^{I,J}\big)
\end{eqnarray}
with the constraints $ \sum_i m_i^I = - \tilde k \frac{d}{N} (\text{ mod }r \frac{d}{N})$.
The $R$-charge of the field $\Phi_{I,J}$ has been denoted as $\Delta_{I,J}$ and
 we also turned on  the chemical potentials $\mu_{\{I,J\}}$ that take into account the possible 
non $R$-global charges for each field.

\bibliographystyle{ytphys}
\bibliography{ref}

\end{document}